\documentclass[conference,letterpaper]{IEEEtran}
\usepackage[utf8]{inputenc}
\usepackage[T1]{fontenc}
\usepackage{url}
\usepackage{cite}
\usepackage[cmex10]{amsmath}
\usepackage{amssymb}
\usepackage{graphicx}
\usepackage[a-2b]{pdfx}
\hypersetup{hidelinks}
\newtheorem{theorem}{Theorem}
\newtheorem{lemma}{Lemma}
\newtheorem{corollary}{Corollary}
\newcommand{\R}{\mathbb{R}}
\newcommand{\E}{\mathbb{E}}
\DeclareMathOperator*{\argmin}{arg\,min}
\newcommand{\rtri}{r_\ast}

\newcommand{\SqBound}{0.7254}
\newcommand{\SqError}{0.7268}
\newcommand{\SqT}{1.3195}
\newcommand{\TriSBeta}{0.80063}
\newcommand{\TriSChiA}{1.00344}
\newcommand{\TriSChiB}{1.61556}
\newcommand{\TriSChiC}{0.42175}
\newcommand{\TriSChiHi}{-0.0298}
\newcommand{\TriSChiLo}{-0.0207}
\newcommand{\TriSChiRootHi}{1.2822}
\newcommand{\TriSChiRootLo}{0.3278}
\newcommand{\TriSEta}{0.65399}
\newcommand{\TriSFa}{0.28318}
\newcommand{\TriSHi}{1.25}
\newcommand{\TriSIntervals}{1750}
\newcommand{\TriSLeft}{0.2586}
\newcommand{\TriSLo}{0.35}
\newcommand{\TriSMargin}{0.0032}

\newcommand{\TriSPanels}{3000}
\newcommand{\TriSPanelsLeft}{4000}
\newcommand{\TriSPrec}{80}
\newcommand{\TriSStep}{10^{-3}}
\newcommand{\TriST}{3}
\newcommand{\TriSTail}{0.2171}
\newcommand{\TriSThetaOne}{0.6745}
\newcommand{\TriSZstar}{1.15399}

\begin{document}
\title{The Optimal Three-Point Quantizer\\ of the Standard Gaussian}

\author{%
  \IEEEauthorblockN{Yutong Gou and Cheng Long}
  \IEEEauthorblockA{College of Computing and Data Science\\
                    Nanyang Technological University, Singapore\\
                    Email: yutong003@e.ntu.edu.sg, c.long@ntu.edu.sg}
}

\maketitle

\begin{abstract}
Optimal quantizers of the standard Gaussian are known in one dimension, and with two points in
every dimension. We prove that, for every $p\ge2$, the optimal three-point quantizer of
$N(0,I_p)$ is unique up to orthogonal transformations: it is the equilateral triangle centered
at the origin with circumradius $3\sqrt3/(2\sqrt{2\pi})$. This settles the planar case of
Tarpey's conjecture that the regular simplex gives the optimal $p+1$ points of the standard
Gaussian. A lower bound on the squared error, in which each codeword contributes a term that
depends only on its norm, reduces the planar problem to an inequality for a function of one
variable. We prove this inequality analytically near the maximizer of the function and for
large arguments. A validated computation (interval arithmetic) of one-dimensional integrals
proves it on the two remaining compact intervals. A projection argument of Flury extends the
result to every dimension $p\ge2$. As a corollary, among all partitions of the plane into three
cells, the three $120^\circ$ sectors uniquely maximize the variance between cells, up to
rotation and null sets. For cells of equal mass this follows from work of Khot and Naor, but
not for unequal masses.
\end{abstract}

\section{Introduction}
\label{sec:intro}

A $K$-point quantizer, or codebook, $C=\{c_1,\dots,c_K\}\subset\R^p$ maps an input $x$ to a
nearest codeword $Q_C(x)\in\argmin_{c\in C}\|x-c\|^2$, where $\|\cdot\|$ is the Euclidean norm.
For $X\sim N(0,I_p)$, we call $\E\|X-Q_C(X)\|^2$ the \emph{squared error} of $C$. The codewords
need not be distinct. A codebook is \emph{optimal} if no codebook with the same number of
points has a smaller squared error. Rotations and reflections leave $N(0,I_p)$, and hence the
squared error, unchanged. So an optimal codebook can be unique only up to orthogonal
transformations.

We call a codebook a \emph{Lloyd fixed point} if every codeword is the centroid of its Voronoi
cell. In one dimension, if the logarithm of the density is concave, there is a unique Lloyd
fixed point with $K$ distinct codewords~\cite{fleischer1964,trushkin1982,kieffer1983}. The
standard Gaussian satisfies this condition because its log-density is $-x^2/2$ plus a constant.
Since every optimal codebook is a Lloyd fixed point~\cite{lloyd1982,max1960,gershogray1992},
this unique fixed point is the optimal $K$-point codebook. In higher dimensions, a Lloyd fixed
point is not necessarily optimal: Gray and Karnin found several locally optimal Gaussian
quantizers in two and three dimensions~\cite{graykarnin1982}. Numerically optimized Gaussian
quantizers are tabulated~\cite{pagesprintems2003}, but proofs are rare. The two-point case is
known in every dimension because it reduces to finding the optimal two-point codebook for a
one-dimensional standard Gaussian~\cite[Theorem~2]{flury1990},
\cite[Example~4.20]{grafluschgy2000}. Beyond the one-dimensional and the two-point case, we are
not aware of any proof of optimality for the standard Gaussian.

For the two-dimensional standard Gaussian, an equilateral triangle centered at the origin was
already a candidate for the optimal three-point codebook. As reported by
Tarpey~\cite{tarpey1998}, Iyengar and Solomon~\cite{iyengar1983} found such a triangle that is
a Lloyd fixed point. Flury~\cite{flury1990} obtained the same configuration by numerically
minimizing the squared error. Tarpey~\cite[Sections~3.3 and~4]{tarpey1998} gave its circumradius
$\rtri=3\sqrt3/(2\sqrt{2\pi})$ and squared error $2-27/(8\pi)$, and described the triangle
as optimal while noting that he had no proof. He also conjectured~\cite[Section~5.1]{tarpey1998} that, in every dimension $p$,
the vertices of a regular simplex centered at the origin are the optimal $p+1$ points of
$N(0,I_p)$; for $p=2$ this is the triangle.

\textbf{Contribution.} We prove that, for every $p\ge2$, the triangle is the unique optimal
three-point codebook of $N(0,I_p)$ up to orthogonal transformations
(Theorem~\ref{thm:triangle}). To our knowledge, this is the first proof of optimality for the
standard Gaussian with three or more points in more than one dimension. The proof has three
parts.
\begin{itemize}
  \item We prove a lower bound on the squared error of any codebook (Lemma~\ref{lem:dual}). It
    depends on the codewords only through their norms.
  \item The bound has a free parameter, a function $\lambda$ of $\|X\|$. For one explicit
    $\lambda$, the bound is tight at the triangle. With this $\lambda$, optimality reduces to
    an inequality for one function $F$ of one variable (Lemma~\ref{lem:scalar}). We prove this
    inequality analytically near the maximizer of $F$, where it is tight, and for large
    arguments. On the two remaining compact intervals the inequality holds with a margin.
    There, a validated computation~\cite{tucker2011} (i.e., interval arithmetic with rigorous
    error bounds) of one-dimensional integrals proves it.
  \item A projection argument of Flury~\cite{flury1990} passes from the plane to every
    dimension $p\ge2$.
\end{itemize}
As a corollary, among all partitions of the plane into three cells, the three $120^\circ$
sectors uniquely maximize the variance between cells, up to rotation and null sets
(Corollary~\ref{cor:partition}). The bound of Lemma~\ref{lem:dual} does not settle the next
cases: Section~\ref{sec:partition} shows that it cannot be tight for the regular tetrahedron,
and that for four points in the plane the lower bound that it gives for all codebooks is
smaller than the squared error of the numerically optimal square. The program of the validated
computation is public.\footnote{\url{https://github.com/gouyt13/gaussian-three-point}}

\textbf{Related results.} Two restricted versions of the problem are covered by known results.
For $p+1$ codewords of equal norm in $\R^p$, the optimality of the regular simplex is
equivalent to the simplex mean width
conjecture~\cite{litvak2018simplex,kabluchko2017meanwidth}. Its three-point case is elementary
(among triangles inscribed in a circle, the equilateral one has the largest perimeter).
According to preprints, the four-point case~\cite{sun2020meanwidth,goldsmith2021meanwidth} and the
general case~\cite{mulgund2026simplex} hold. For three cells of equal
mass, the optimality of the triangle follows from a theorem of Khot and
Naor~\cite{khot2009kernel} on partitions of the plane (Section~\ref{sec:partition}). Their
propeller conjecture extends this theorem to every dimension and every number of cells; it was
proved in $\R^3$~\cite{heilman2013propeller} and, according to recent manuscripts, in every
dimension~\cite{openai2026propeller,heilman2026propeller}. A general codebook need not have
codewords of equal norm or cells of equal mass.

\section{A Dual Bound}
\label{sec:dual}

Let $X\sim N(0,I_p)$, $R=\|X\|$, and $e_1=(1,0,\dots,0)$. Let the function $\lambda$ be
measurable on $[0,\infty)$ with $\E|\lambda(R)|<\infty$. For $t\ge0$ define
\[
W_\lambda(t)=\E\big(\lambda(R)-\|X-te_1\|^2\big)_+ ,\quad (y)_+=\max(y,0).
\]
This is the expected amount by which the squared distance from $X$ to a point of norm $t$ falls
below the threshold $\lambda(R)$.

\begin{lemma}[Dual bound]\label{lem:dual}
For every codebook $C$ of $K$ points $c_1,\dots,c_K$ in $\R^p$,
\[
\E\|X-Q_C(X)\|^2\ \ge\ \E\lambda(R)-\sum_{i=1}^KW_\lambda(\|c_i\|).
\]
\end{lemma}
\begin{IEEEproof}
Let $d=\|X-Q_C(X)\|^2=\min_i\|X-c_i\|^2$. Then
\begin{align*}
d&\ge\lambda(R)-(\lambda(R)-d)_+ ,\\
(\lambda(R)-d)_+&=\max_i\big(\lambda(R)-\|X-c_i\|^2\big)_+\\
&\le\sum_{i=1}^K\big(\lambda(R)-\|X-c_i\|^2\big)_+ .
\end{align*}
Take expectations. An orthogonal map that sends $c_i$ to $\|c_i\|e_1$ leaves the distribution
of $X$ and the value of $R$ unchanged, so the $i$th term has expectation $W_\lambda(\|c_i\|)$.
\end{IEEEproof}

Lemma~\ref{lem:dual} is weak Lagrangian duality. For a fixed codebook, the squared error is the
least cost of assigning every input to exactly one codeword. Replace the constraint ``exactly
one'' by a multiplier $\lambda(\|x\|)$ at each input $x$: the cost of $x$ becomes
$\lambda(\|x\|)$, plus $\|x-c_i\|^2-\lambda(\|x\|)$ for each codeword $c_i$ that takes $x$. The
cheapest choice gives $x$ to every $c_i$ with $\|x-c_i\|^2<\lambda(\|x\|)$, whatever the other
codewords are, and its expected cost is the right side of Lemma~\ref{lem:dual}. Bounding each
term by $\sup_tW_\lambda(t)$ gives one lower bound for all codebooks of $K$ points. This bound
has the form of the dual of the $k$-median linear programming relaxation for finite point
sets~\cite{awasthi2015relax}, with squared distances in place of distances.
In~\cite{awasthi2015relax}, dual solutions of this kind certify optimal clusterings.

Equality holds in Lemma~\ref{lem:dual} exactly when both inequalities in the proof are
equalities almost surely. The first is an equality exactly when $\lambda(R)\ge d$, and the
second exactly when at most one term of the sum is positive. So, for $K\ge2$, equality holds
exactly when, for almost every $x$, the value $\lambda(\|x\|)$ lies between the squared
distances from $x$ to its nearest and to its second-nearest codeword, where copies of a
repeated codeword count separately. In particular, the sets $\{x:\lambda(\|x\|)>\|x-c_i\|^2\}$
are disjoint up to null sets. A point on the boundary between two cells has the same distance
to its two nearest codewords, and both distances are continuous in $x$. By Fubini's theorem in polar coordinates, for almost every $r$, the equality condition holds at almost every point of norm $r$, hence on a dense subset of the sphere of radius $r$. Thus, for almost every
$r$ for which a point of norm $r$ lies on such a boundary, $\lambda(r)$ must be the squared
distance from this point to its nearest codewords.

\section{The Three-Point Theorem}
\label{sec:theorem}

\begin{figure}[t]
\centering
\includegraphics{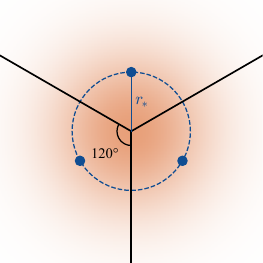}
\caption{The optimal three-point codebook of $N(0,I_2)$ (Theorem~\ref{thm:triangle}). The
shading shows the density. The codewords lie on the dashed circle of radius $\rtri$.
}
\label{fig:triangle}
\end{figure}

\begin{theorem}\label{thm:triangle}
Let $p\ge2$ and $X\sim N(0,I_p)$. Up to orthogonal transformations, the unique optimal
three-point quantizer of $X$ is the equilateral triangle centered at the origin with
circumradius
\[
\rtri=\frac{3\sqrt3}{2\sqrt{2\pi}}=1.0364824\ldots
\]
in a plane through the origin. Its squared error is $p-27/(8\pi)$.
\end{theorem}
Fig.~\ref{fig:triangle} shows the triangle.

In outline, the proof is as follows. For $p=2$, Lemma~\ref{lem:dual} with one explicit
$\lambda$ bounds the squared error of a codebook $\{c_1,c_2,c_3\}$ from below by
$\E\lambda(R)-\sum_iF(\|c_i\|)$, where $F$ is a function of one variable.
Lemma~\ref{lem:scalar} below states that $F$ is largest exactly at $\rtri$, so this bound is at
least $2-27/(8\pi)$, the squared error of the triangle. The conditions for equality then give
uniqueness, and a projection argument extends the result to $p>2$.

Let $p=2$. We apply Lemma~\ref{lem:dual} with
\[
\lambda(r)=r^2-\rtri r+\rtri^2 ,
\]
the squared distance between points of norms $r$ and $\rtri$ at angle $\pi/3$. A point on the
boundary between two cells of the triangle is at angle $\pi/3$ from its two nearest vertices,
so this is the $\lambda$ that the equality condition of Section~\ref{sec:dual} requires. For
this $\lambda$, expanding $\|X-te_1\|^2$ with $X_1=\langle X,e_1\rangle$ gives
\[
\lambda(R)-\|X-te_1\|^2=2tX_1-\rtri R+\rtri^2-t^2 .
\]
So $W_\lambda(t)=F(t)$, where
\[
F(t,\tau)=\E\big(2tX_1-\rtri R+\rtri^2-\tau\big)_+ ,\quad F(t)=F(t,t^2).
\]
The second argument $\tau$ replaces $t^2$; it is used only in Section~\ref{sec:scalar}. Let
$\varphi$ be the polar angle of $X$ measured from $e_1$, so $X_1=R\cos\varphi$. In polar
coordinates, the term in brackets is $R\,(2t\cos\varphi-\rtri)+\rtri^2-\tau$. Since $p=2$, the
angle $\varphi$ is uniform and independent of $R$, and $R$ has the Rayleigh density
$re^{-r^2/2}$. We write $\mu=\E R=\sqrt{\pi/2}$; also $\E R^{-1}=\mu$ and $\rtri\mu=3\sqrt3/4$.

\begin{lemma}[Scalar inequality]\label{lem:scalar}
For all $t\ge0$,
\[
F(t)\le F(\rtri)=\frac{9}{4\pi}-\frac{\sqrt3}{4},
\]
with equality only for $t=\rtri$.
\end{lemma}

\begin{figure}[t]
\centering
\includegraphics[width=\columnwidth]{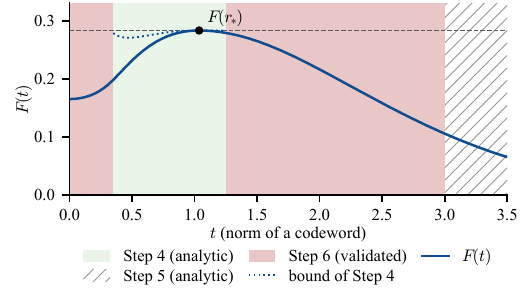}
\caption{The function $F$ of Lemma~\ref{lem:scalar} (floating-point evaluation, for
illustration) and the intervals covered by Steps~4--6 of the proof. The dashed line is
$F(\rtri)$. The dotted curve is the function $G$ of Step~4, drawn on $[\TriSLo,\TriSHi]$: $F\le G$, with equality at $\rtri$.}
\label{fig:scalar}
\end{figure}

Section~\ref{sec:scalar} proves the lemma, with a validated computation in Step~6, and Fig.~\ref{fig:scalar} shows $F$ and which step
of the proof covers which interval.

Next, we prove Theorem~\ref{thm:triangle} using Lemma~\ref{lem:scalar}.

\begin{IEEEproof}[Proof of Theorem~\ref{thm:triangle}]
\emph{The lower bound for $p=2$.} Let $C=\{c_1,c_2,c_3\}$ be any codebook. Since $\E R^2=2$,
\[
\E\lambda(R)=2-\rtri\mu+\rtri^2=2-\frac{3\sqrt3}{4}+\frac{27}{8\pi}.
\]
By Lemmas~\ref{lem:dual} and~\ref{lem:scalar},
\begin{align}
\E\|X-Q_C(X)\|^2&\ge\E\lambda(R)-\sum_{i=1}^3F(\|c_i\|)\notag\\
&\ge\E\lambda(R)-3F(\rtri)=2-\frac{27}{8\pi}.\label{eq:lower}
\end{align}

\emph{The triangle attains the bound.} Let $C$ now be the triangle with vertices
$c_1,c_2,c_3$ of norm $\rtri$ at mutual angles $2\pi/3$. The Voronoi cell of $c_j$ is the
sector $S_j$ of opening angle $2\pi/3$ around $c_j$. In polar coordinates $(R,\varphi)$ with
$\varphi$ measured from $c_j$,
\[
\E\big[\langle X,c_j\rangle\mathbf 1\{X\in S_j\}\big]
=\frac{\rtri\mu}{2\pi}\int_{-\pi/3}^{\pi/3}\cos\varphi\,d\varphi=\frac{\sqrt3\,\rtri\mu}{2\pi}.
\]
Since $Q_C(X)=c_j$ on $S_j$, summing over $j$ gives
$\E\langle X,Q_C(X)\rangle=3\sqrt3\,\rtri\mu/(2\pi)=27/(8\pi)$. Also
$\|Q_C(X)\|^2=\rtri^2=27/(8\pi)$, so the squared error is
$2-2\cdot27/(8\pi)+27/(8\pi)=2-27/(8\pi)$.

\emph{Uniqueness for $p=2$.} Suppose that equality holds in~\eqref{eq:lower} for a codebook
$C=\{c_1,c_2,c_3\}$. Then both inequalities there are equalities: equality holds in
Lemma~\ref{lem:dual}, and $F(\|c_i\|)=F(\rtri)$ for every $i$. Lemma~\ref{lem:scalar} then
gives $\|c_i\|=\rtri$. For a point $c$ of norm $\rtri$,
$\lambda(\|x\|)-\|x-c\|^2=\rtri\|x\|\,(2\cos\varphi-1)$, where $\varphi$ is the angle between
$x$ and $c$. So $\lambda(\|x\|)>\|x-c\|^2$ exactly on the open sector of opening angle $2\pi/3$
around $c$. By Section~\ref{sec:dual}, equality in Lemma~\ref{lem:dual} implies that the three
sectors around $c_1,c_2,c_3$ are disjoint up to null sets. Hence the directions of any two
codewords are at least $2\pi/3$ apart. The three gaps between consecutive directions sum to
$2\pi$, so each equals $2\pi/3$, and $C$ is the triangle.

\emph{Every dimension $p>2$.} We follow the proof of Theorem~3 of Flury~\cite{flury1990}. Let
$C$ be any three-point codebook. Let $V$ be the linear span of the differences of its
codewords, so $\dim V\le2$, and let $\Pi$ be the orthogonal projection onto $V$. The vector
$w=(I-\Pi)c$ is the same for all $c\in C$. Since
$\|x-c\|^2=\|\Pi x-\Pi c\|^2+\|(I-\Pi)x-w\|^2$, a codeword $c$ is nearest to $x$ in $C$ exactly
when $\Pi c$ is nearest to $x$ in $\Pi C$, and $\E X=0$ gives
\[
\E\min_{c\in C}\|X-c\|^2=\E\min_{c\in C}\|X-\Pi c\|^2+\|w\|^2 .
\]
So projecting $C$ onto $V$ lowers the squared error by $\|w\|^2$, which is positive unless
$C\subset V$. Choose a plane $T\supseteq V$ through the origin. The projection of $X$ onto $T$
is standard Gaussian in $T$. The squared error of $\Pi C$ is $p-2$ plus its squared error in
$T$ for this projection. By the case $p=2$, this sum is at least $p-27/(8\pi)$, with equality
exactly when $\Pi C$ is the centered equilateral triangle of circumradius $\rtri$ in $T$. Hence
$C$ has squared error at least $p-27/(8\pi)$. Equality holds exactly when $C\subset V$ and
$\Pi C$ is this triangle, that is, when $C$ is this triangle.
\end{IEEEproof}

\section{Proof of the Scalar Inequality}
\label{sec:scalar}

The proof of Lemma~\ref{lem:scalar} has six steps. Step~1 computes $F(\rtri)$. Steps~2--4 prove
the lemma on $[\TriSLo,\TriSHi]$. Step~2 integrates over the angle and writes $F$ through a
function of one variable. Step~3 bounds this function from above by a parabola. Step~4 computes
the resulting bound on $F$ in closed form. Step~5 proves the lemma for $t\ge\TriST$ by a
Gaussian tail bound. Step~6 proves it on the two remaining intervals, $[0,\TriSLo]$ and
$[\TriSHi,\TriST]$, by a validated computation. Steps~1--5 are analytic. They use only a few
numerical inequalities between closed-form constants.

\emph{Step 1: the value at $\rtri$.} In polar coordinates,
\begin{align*}
F(\rtri)&=\rtri\,\E\big[R\,(2\cos\varphi-1)_+\big]\\
&=\rtri\mu\Big(\frac{\sqrt3}{\pi}-\frac13\Big)=\frac{9}{4\pi}-\frac{\sqrt3}{4}.
\end{align*}

\emph{Step 2: a formula for $F$.} Let $t>0$. For $\|X\|=r>0$, the term in brackets is
\[
2tr\cos\varphi-\rtri r+\rtri^2-t^2=2rt\,\big(\cos\varphi-z_t(r)\big)
\]
with $z_t(r)=(\rtri r+t^2-\rtri^2)/(2rt)$. We take the factor $2rt>0$ out of $(\cdot)_+$. Since
$\varphi$ is uniform and independent of $R$, and $\cos$ is even, integrating over the angle
first gives
\begin{equation}\label{eq:F-polar}
F(t)=\frac1\pi\,\E\big[2Rt\,\psi(z_t(R))\big],
\end{equation}
where $\psi(z)=\int_0^\pi(\cos\varphi-z)_+\,d\varphi$. We have $\psi(z)=0$ for $z\ge1$,
$\psi(z)=-\pi z$ for $z\le-1$, and $\psi(\cos\theta)=\sin\theta-\theta\cos\theta$ for
$\theta\in[0,\pi]$.

\emph{Step 3: a quadratic majorant of $\psi$.} We bound $\psi$ from above by a parabola
$\beta(z_\ast-z)^2$, which has a double root at $z_\ast$. We choose $\beta$ and $z_\ast$ so
that it touches $\psi$ at $z=\frac12$: the two functions and their derivatives agree there.
Since $\psi'(z)=-\arccos z$ on $[-1,1]$, with $\eta=z_\ast-\frac12$ these two conditions read
\[
\beta\eta^2=\frac{\sqrt3}{2}-\frac\pi6,\qquad 2\beta\eta=\frac\pi3 .
\]
Dividing the first by the second gives $\eta=3\sqrt3/\pi-1\approx\TriSEta$. Hence
$\beta=\pi/(6\eta)\approx\TriSBeta$ and $z_\ast\approx\TriSZstar$. We claim that
\begin{equation}\label{eq:major}
\psi(z)\le\beta(z_\ast-z)^2\qquad\text{for all }z\in\R .
\end{equation}
We check~\eqref{eq:major} on three ranges of $z$. For $z\ge1$, \eqref{eq:major} is trivial. For
$z\le-1$, the derivative of $\beta(z_\ast-z)^2+\pi z$ is $\pi-2\beta(z_\ast-z)$. It increases
in $z$ and equals $\frac{2\pi}{3}-3\beta<0$ at $z=-1$, so the function decreases on
$(-\infty,-1]$. At $z=-1$ the function equals $\beta(z_\ast+1)^2-\pi>0$, so it is positive for
all $z\le-1$, which is~\eqref{eq:major} there. For $z=\cos\theta$ with
\mbox{$\theta\in[0,\pi]$}, let
$f(\theta)=\beta(z_\ast-\cos\theta)^2-\sin\theta+\theta\cos\theta$. Then
$f'(\theta)=\sin\theta\,\varepsilon(\theta)$ with
$\varepsilon(\theta)=2\beta(z_\ast-\cos\theta)-\theta$ and
\mbox{$\varepsilon'(\theta)=2\beta\sin\theta-1$}. So $\varepsilon$ decreases on $[0,\theta_1]$,
increases on $[\theta_1,\pi-\theta_1]$ and decreases on $[\pi-\theta_1,\pi]$, where
$\theta_1=\arcsin\frac{1}{2\beta}\approx\TriSThetaOne<\frac\pi3<\pi-\theta_1$. Moreover
$\varepsilon(0)=\frac\pi3-\beta>0$, $\varepsilon(\frac\pi3)=2\beta\eta-\frac\pi3=0$ and
$\varepsilon(\pi)=3\beta-\frac{2\pi}{3}>0$. Since $\varepsilon$ increases on
$[\theta_1,\pi-\theta_1]$ and vanishes at $\frac\pi3$, it is negative on $[\theta_1,\frac\pi3)$
and positive on $(\frac\pi3,\pi-\theta_1]$. On $[0,\theta_1]$ it decreases from
$\varepsilon(0)>0$ to $\varepsilon(\theta_1)<0$, so it has exactly one zero $\theta_0$ there.
On $[\pi-\theta_1,\pi]$ it decreases to $\varepsilon(\pi)>0$, so it is positive. As
$\sin\theta\ge0$, $f'$ has the sign of $\varepsilon$, so $f$ increases on $[0,\theta_0]$,
decreases on $[\theta_0,\frac\pi3]$ and increases on $[\frac\pi3,\pi]$. Since
$f(0)=\beta(z_\ast-1)^2>0$ and
$f(\frac\pi3)=\beta\eta^2-\big(\frac{\sqrt3}{2}-\frac\pi6\big)=0$, we get $f\ge0$ on $[0,\pi]$,
which proves~\eqref{eq:major}.

\emph{Step 4: the interval $[\TriSLo,\TriSHi]$.} We bound $F$ by an explicit function $G$ and
compare $G$ with $F(\rtri)$. By~\eqref{eq:F-polar} and~\eqref{eq:major}, $F(t)\le G(t)$, where
\[
G(t)=\frac{\beta}{\pi}\,\E\big[2Rt\,(z_\ast-z_t(R))^2\big].
\]
Since $2Rt\,z_t(R)=\rtri R+t^2-\rtri^2$, we have $2Rt\,(z_\ast-z_t(R))=uR-v$ with
$u=2z_\ast t-\rtri$ and $v=t^2-\rtri^2$. Since $\E R=\E R^{-1}=\mu$,
\[
G(t)=\frac{\beta}{2\pi t}\,\E\,\frac{(uR-v)^2}{R}=\frac{\beta}{2\pi t}\big(\mu(u^2+v^2)-2uv\big).
\]
We expand $u$ and $v$ and use $2\beta\eta=\pi/3$, $F(\rtri)=\frac13\eta\mu\rtri$ from Step~1,
and $\mu(1+\eta)=2\rtri$. This gives
\begin{equation}\label{eq:central}
G(t)=F(\rtri)+\frac{(t-\rtri)^2}{2\pi t}\,\chi(t),
\end{equation}
where $\chi(t)=\beta\mu\big(1+(t+\rtri)^2\big)-2\beta(t+\rtri)-\frac{2\pi}{3}t$. In particular
$G(\rtri)=F(\rtri)$: at $t=\rtri$, $z_t(R)=\frac12$ for every $R$, where the two sides
of~\eqref{eq:major} agree. The function $\chi$ is a convex quadratic,
$\chi(t)\approx\TriSChiA\,t^2-\TriSChiB\,t+\TriSChiC$. Its roots are approximately
$\TriSChiRootLo$ and $\TriSChiRootHi$, and we take the interval slightly inside them:
$\chi(\TriSLo)<\TriSChiLo$ and $\chi(\TriSHi)<\TriSChiHi$. Since $\chi$ is convex, $\chi<0$ on
$[\TriSLo,\TriSHi]$. So $F(t)\le G(t)<F(\rtri)$ there unless $t=\rtri$.

\emph{Step 5: large arguments.} We bound $F(t)$ by an explicit function of $t$ that decreases
in $t$, and check that this function is below $F(\rtri)$ at $t=\TriST$. Let $t\ge\TriST$. Since
$R\ge X_1$ and $\rtri>0$, the term in brackets is at most $(2t-\rtri)X_1-(t^2-\rtri^2)$. Hence
\[
F(t)\le(2t-\rtri)\,\E(X_1-r_0)_+,\qquad r_0=\frac{t^2-\rtri^2}{2t-\rtri}>0 .
\]
The coordinate $X_1$ is standard normal, with density $g(x)=e^{-x^2/2}/\sqrt{2\pi}$, and
$g'(x)=-x\,g(x)$. Let $s>0$. For $x\ge s$,
\[
\Pr(X_1>x)\le\int_x^\infty\frac yx\,g(y)\,dy=\frac{g(x)}{x}\le\frac{x\,g(x)}{s^2}.
\]
Hence
\begin{align*}
\E(X_1-s)_+&=\int_s^\infty\Pr(X_1>x)\,dx\\
&\le\int_s^\infty\frac{x\,g(x)}{s^2}\,dx=\frac{g(s)}{s^2}.
\end{align*}
With $s=r_0$,
\[
F(t)\le g(r_0)\,\frac{(2t-\rtri)^3}{(t^2-\rtri^2)^2}.
\]
Both factors are positive and decrease in $t>\rtri$, so the bound decreases in $t$. The first
factor decreases because $g$ decreases on $[0,\infty)$ and
\[
\frac{dr_0}{dt}=\frac{2(t^2-\rtri t+\rtri^2)}{(2t-\rtri)^2}>0 .
\]
The second factor decreases because its logarithmic derivative is
\[
-\frac{2\big((t-\rtri)^2+2\rtri^2\big)}{(2t-\rtri)(t^2-\rtri^2)}<0.
\]
At $t=\TriST$ the bound is below $\TriSTail$, while $F(\rtri)>\TriSFa$. So $F(t)<F(\rtri)$ for
all $t\ge\TriST$.

\emph{Step 6: the intervals $[0,\TriSLo]$ and $[\TriSHi,\TriST]$.} We split each interval into
subintervals $[t_1,t_2]$. First, we bound $F$ on each subinterval by one number. Second, we
bound this number from above by a finite sum. Third, we evaluate the sum in interval
arithmetic.

First, for fixed $\tau$, the function $F(t,\tau)$ is convex in $t$, as the mean of the positive part of an affine function of $t$. It is even in $t$, because
$(X_1,R)$ and $(-X_1,R)$ have the same distribution. So it is nondecreasing in $t\ge0$. It is
nonincreasing in $\tau$. Hence
\begin{equation}\label{eq:sandwich}
F(t)=F(t,t^2)\le F(t_2,t_1^2)\qquad\text{for }0\le t_1\le t\le t_2 .
\end{equation}

Second, in polar coordinates, $F(t,\tau)=\frac1\pi\int_0^\pi I(\varphi)\,d\varphi$ with
$I(\varphi)=\E_R(Rk-\ell)_+$, where $k=2t\cos\varphi-\rtri$ and $\ell=\tau-\rtri^2$. Since
$(Rk-\ell)_+$ is nondecreasing in $k$ and $k$ decreases in $\varphi\in[0,\pi]$, $I$ is
nonincreasing in $\varphi$. Hence
$\frac1\pi\int_0^\pi I(\varphi)\,d\varphi\le\frac1N\sum_{j=0}^{N-1}I(j\pi/N)$ for every $N$. To
evaluate $I(j\pi/N)$, we express $I$ through $m(s)=\E(R-s)_+$; the angle enters only through
$k$. For $s\ge0$,
$m(s)=\int_s^\infty\Pr(R>x)\,dx=\int_s^\infty e^{-x^2/2}\,dx=\mu\,\mathrm{erfc}(s/\sqrt2)$. For
$s<0$, $m(s)=\mu-s$, since $R\ge0$. For $k>0$, $(Rk-\ell)_+=k\,(R-\ell/k)_+$, so
$I=k\,m(\ell/k)$. For $k<0$, the identity $y_+=y+(-y)_+$ gives
$(Rk-\ell)_+=Rk-\ell-k\,(R-\ell/k)_+$, so $I=k\mu-\ell-k\,m(\ell/k)$.

Third, we evaluate the sums in the ball arithmetic (midpoint--radius interval arithmetic) of
Arb~\cite{johansson2017arb} at $\TriSPrec$ bits. For every arithmetic operation, for $\pi$, and
for $\sqrt{\cdot}$, $\cos$, $\arcsin$, $\exp$ and $\mathrm{erfc}$, Arb returns a ball that
contains the exact result. The sign of $\ell$ is fixed on each interval, and no ball for $k$
contains zero, so we know which formula for $I$, and which formula for $m$, applies to each
term.

On $[0,\TriSLo]$, one subinterval and $N=\TriSPanelsLeft$ give
$F(\TriSLo,0)<\TriSLeft<F(\rtri)$. On $[\TriSHi,\TriST]$, \TriSIntervals{} subintervals of
length $\TriSStep$ and $N=\TriSPanels$ give $F(t_2,t_1^2)<F(\rtri)-\TriSMargin$ on each.
By~\eqref{eq:sandwich}, $F(t)<F(\rtri)$ on both intervals. The same program also checks the
numerical inequalities used in Steps~3--5. It takes 12--14 seconds on one core and
rests on the correctness of Arb and of a short program. Steps~4, 5 and~6 together prove the
lemma. \hfill\IEEEQEDclosed

\section{Partitions and Outlook}
\label{sec:partition}

Theorem~\ref{thm:triangle} also answers a question about partitions of the plane. For $p=2$ and
a partition $\mathcal A=(A_1,\dots,A_K)$ of $\R^2$ into measurable cells, let
$P_i=\Pr(X\in A_i)$ and $M_i=\E[X\,\mathbf 1\{X\in A_i\}]$, and define
$\Lambda(\mathcal A)=\sum_{i:\,P_i>0}\|M_i\|^2/P_i$. For a cell with $P_i>0$ and any point $c$,
\begin{align*}
\E\big[\|X-c\|^2\mathbf 1\{X\in A_i\}\big]
&=\E\big[\|X\|^2\mathbf 1\{X\in A_i\}\big]-\frac{\|M_i\|^2}{P_i}\\
&\quad+P_i\Big\|c-\frac{M_i}{P_i}\Big\|^2 .
\end{align*}
So the map that sends each cell $A_i$ to its centroid $M_i/P_i$ has squared error
$2-\Lambda(\mathcal A)$. Since $\E X=0$, $\Lambda(\mathcal A)$ is the variance between cells:
each cell contributes its mass times the squared norm of its centroid.

\begin{corollary}\label{cor:partition}
Every partition of $\R^2$ into three measurable cells satisfies $\Lambda\le27/(8\pi)$. Equality
holds exactly for the three $120^\circ$ sectors from the origin, up to rotation and null sets.
\end{corollary}
\begin{IEEEproof}
The sectors attain the bound: their centroids are the vertices of the triangle, so $2-\Lambda$
is the squared error of the triangle. Conversely, take the centroids of the cells of positive
mass as a codebook, and repeat one of them if there are fewer than three. Its nearest-codeword
map is pointwise at least as good as the map that sends each cell to its centroid, so its
squared error is at most $2-\Lambda$. By Theorem~\ref{thm:triangle}, which allows repeated
codewords, this squared error is at least $2-27/(8\pi)$. Hence $\Lambda\le27/(8\pi)$. If
equality holds, both maps have squared error $2-27/(8\pi)$. So the codebook is optimal, hence
the triangle, and all three cells have positive mass. Moreover, the centroid map sends almost
every input to a nearest codeword, so the cells are the Voronoi cells of the triangle up to
null sets.
\end{IEEEproof}

Khot and Naor~\cite{khot2009kernel} proved that, over partitions of the plane into three cells,
the three sectors maximize $\sum_i\|M_i\|^2$, which is $\Lambda$ without the weights $1/P_i$.
For cells of mass $1/3$, $\Lambda=3\sum_i\|M_i\|^2$, so the bound of
Corollary~\ref{cor:partition} follows from their theorem, and its equality case from their
proof, as noted in~\cite[Section~4]{heilman2026propeller}. For unequal masses neither follows.

\textbf{Outlook.} Lemma~\ref{lem:dual} holds for every number of points and every $\lambda$.
The proof of Theorem~\ref{thm:triangle} needs two properties of $\lambda$: the bound is tight
at the candidate codebook, and $W_\lambda$ is largest at the norm of its codewords
(Lemma~\ref{lem:scalar}). By Section~\ref{sec:dual}, the first requires $\lambda(\|x\|)$ to lie
between the squared distances from $x$ to its nearest and to its second-nearest codeword, for
almost every $x$. The next two cases each fail one property.

The next case of Tarpey's conjecture, the regular tetrahedron in $\R^3$, fails the first. Let
its vertices be $c_1,\dots,c_4$, where $\|c_i\|=\rho>0$ and
$\langle c_i,c_j\rangle=-\rho^2/3$ for $i\ne j$. At norm $r>0$, the squared distance to the nearest
vertex is $r^2+\rho^2-2r\rho/3$ in the direction $-c_1$, where $c_2$, $c_3$ and $c_4$ are equally near. The squared distance to the
second-nearest vertex is $r^2+\rho^2-2r\rho/\sqrt3$ in the direction $c_1+c_2$, where $c_1$ and $c_2$ are equally near. So the equality condition holds in both directions only if $\lambda(r)$ is at least the first value and at most the second. The second
value is smaller, so no value of $\lambda(r)$ lies between the two distances in both
directions. Both distances are continuous in the direction, so the same holds on sets of
positive measure, and no $\lambda$ makes the bound tight.

The square, the numerically optimal four points in the plane~\cite{flury1990}, fails the
second. As a Lloyd fixed point, it has circumradius $\rho=2/\sqrt\pi$ and squared error
$2-4/\pi\approx\SqError$. The first property forces, for almost every $r$,
$\lambda(r)=r^2-\sqrt2\rho r+\rho^2$, the squared distance from a boundary point of norm $r$ to
its two nearest vertices. Then $\E\lambda(R)-4W_\lambda(\rho)=2-4/\pi$. But a validated
computation gives $W_\lambda(\SqT)>W_\lambda(\rho)$, so the bound
$\E\lambda(R)-4\sup_tW_\lambda(t)$, about $\SqBound$, is below $2-4/\pi$ and does not prove
that the square is optimal. For every $\lambda$ without the first property, $\E\lambda(R)-4\sup_tW_\lambda(t)\le\E\lambda(R)-4W_\lambda(\rho)<2-4/\pi$ as well.

\newpage
\section*{Acknowledgment}
\textbf{Use of generative AI.} Claude models of Anthropic (Claude Opus~5.5, Claude Fable~5.1
and Claude Fable~5.2), through Claude Code, were used in this work; GPT-6 Astra of OpenAI,
through Codex, was used in early discussions. The AI systems proposed the proof strategy,
drafted the proofs in Sections~II--V, wrote the program of the validated computation in Step~6,
and drafted the text of all sections. The first author selected the problem, chose the proof
strategy among the candidates, directed the work, and decided which results to claim. Both
authors revised the text, checked every step of the proofs, and take full responsibility for
the content.

\IEEEtriggeratref{10}
\bibliographystyle{IEEEtran}
\bibliography{refs}

\end{document}